\documentclass[12pt]{article}
\usepackage{amscd}
\usepackage{bbm}
\usepackage{mathrsfs}
\usepackage{amssymb}
\usepackage{graphics}
\usepackage{graphicx}
\usepackage{amsfonts}
\usepackage{amsmath}
\usepackage{float}
\usepackage[numbers,sort&compress]{natbib}

\begin{document}
\date{}
\title{The orbital stability of the periodic traveling wave solutions to the defocusing complex modified Korteweg-de Vries equation}
\author{ Wen-Rong Sun \thanks{swrustb@163.com} \\ 
School of Mathematics and Physics,\\ University
of Science and Technology Beijing, Beijing 100083, China}
\maketitle \vspace{-0.8cm}

%\newpage

%\newpage
\begin{abstract}
The stability of the elliptic solutions to the defocusing complex modified Korteweg-de Vries (cmKdV) equation is studied. Using the integrability of the defocusing cmKdV equation, we prove the spectral stability of the elliptic solutions.  We show that one special linear combination of the first five conserved quantities
produces a Lyapunov functional, which implies that the elliptic solutions are orbitally stable with respect to the subharmonic perturbations.
\textbf{AMS subject classications.} 37K45, 35Q55, 33E05
\end{abstract}
\newpage

\noindent\textbf{\large 1. Introduction}\\\hspace*{\parindent}
The complex modified Korteweg-de Vries (cmKdV) equation~\cite{sinh1}
\begin{eqnarray}\label{cmkdv}
u_{\mathrm{t}}+6\sigma|u|^{2} u_{x}+u_{x x x}=0, \quad \sigma=\pm1,
\end{eqnarray}
is the third flow of the nonlinear Schr\"{o}dinger (NLS) equation hierarchy. 
When $\sigma=-1$, (\ref{cmkdv}) corresponds to the defocusing case and (\ref{cmkdv}) corresponds to the focusing case when $\sigma=1$.
In addition to the NLS equation, as stated in~\cite{zrf1}, (\ref{cmkdv}) also have a universal character. Due to the presence of the third-order dispersion term, (\ref{cmkdv}) is relevant for ultrashort
pulse transmission~\cite{yjk}. Many aspects of (\ref{cmkdv}) have been investigated. For example,  the inverse scattering transformation of~(\ref{cmkdv}) has been investigated~\cite{fss1} and the authors in~\cite{sdd1} have studied the Whitham equations for the defocusing case of (\ref{cmkdv}). Using Kato's theory, the local well-posedness for (\ref{cmkdv}) in the non-periodic case has been studied in~\cite{rp}.  In~\cite{rp2}, the authors have proved that in periodic Sobolev  spaces $H^{s}$, the problem is locally well-posed for $s>\frac{3}{2}$.

The stability of solitary waves to nonlinear dispersive equations was first studied by Benjamin~\cite{bt}.
Cazenave and Lions proved the orbital stability of the soliton in 1982~\cite{6211}. Later, Weinstein proved the orbital stability of the soliton using Lyapunov techniques~\cite{lf1}.
Maddocks and Sachs proved the stability of multi-soliton solutions to the KdV equation~\cite{by15} and this classical work was generalized to a larger class of integrable systems by Kapitula~\cite{6212}.
Using regular perturbation theory and  treating the Floquet parameter as a small parameter, Rowlands~\cite{6213} was the first to study the stability of the stationary
periodic solutions to the focusing NLS equation. Using the energy method, Gallay and Haragus proved the fact that the periodic waves are orbitally stable
within a class of solutions which have the same periodicity properties as the wave itself~\cite{6214,6215}. Later, Gallay and Pelinovsky proved the orbital stability of elliptic solutions to the defocusing NLS equation with respect to subharmonic perturbations~\cite{TG}.   
Haragus and Kapitula~\cite{by13} considered the problem of determining the spectrum for the linearization of an infinite-dimensional Hamiltonian system about a spatially periodic traveling wave.  They established the spectral instability for the quasi-periodic solutions of sufficiently small amplitude. 
Using tools from integrability theory, Deconinck and their collaborators have proved the spectral stability and orbital stability of many integrable systems,
such as the KdV equation~\cite{bd8,655}, the defocusing NLS equation~\cite{bd6}, the focusing NLS equation~\cite{t1,by90}, the modified KdV equation~\cite{bd5}, the sine-Gordon equation~\cite{ees1} and the sinh-Gordon equation~\cite{swr}. In 2020, Deconinck and Upsal ~\cite{by90} examined the stability of the elliptic solutions of the focusing NLS equation with respect to subharmonic perturbations and that work is the first in the program to establish the orbital stability for elliptic solutions for which the underlying Lax pair is not self-adjoint using the integrable method.  

The motivations of our paper are given below.

1. In~\cite{rp2}, the authors have studied the existence and stability of the periodic traveling-wave solutions to the focusing cmKdV equation using the PDE techniques. In Section 3 of~\cite{rp2}, the authors claimed that they can't handle the stability analysis of the defocusing cmKdV equation since the periodic solutions are given by the periodic trajectories of the Hamiltonian vector field which oscillate around the center at the origin in the phase portraits. Our aim in this paper is to deal with such case using the integrability of the defocusing cmKdV equation.
Besides, we consider the subharmonic perturbations which are periodic perturbations having period equal to an integer multiple of the period of the potential solution.  In fact, for many integrable PDEs, such as the NLS equation~\cite{t1, by90}, the modified KdV equation~\cite{bd5}, and the sine-Gordon equation~\cite{ees1},  the elliptic solutions are stable with respect to coperiodic perturbations, but unstable with respect to subharmonic perturbations~\cite{bd5,t1,ees1}.  Therefore, we consider the perturbations which are not limited to coperiodic perturbations~\cite{TG,swr,bd5,bd6,bd7,bd8,t1,jm}. 

2.  A classical tool for numerically computing the Lax spectrum for periodic potentials is the Floquet discriminant~\cite{6222,6223,6224}. But, one couldn’t give a full description of the Lax spectrum analytically using the Floquet discriminant. Ivey and Lafortune established a connection between the Floquet spectrum and the stability properties of the solutions to the Hirota equation~\cite{6224}. Different from the Floquet discriminant, we determine the linear stability and nonlinear stability from the Lax spectral problem analytically.

Based on the above motivations, we aim to study the stability of elliptic solutions to the defocusing cmKdV equation with respect to subharmonic perturbations. 
Now, we briefly introduce the integrable method~\cite{bd8,655,bd6,t1,by90,bd5} used in this paper.  For the spectral stability, we need to determine the Lax spectrum firstly. Then we construct the squared-eigenfunction connection for the defocusing cmKdV equation, which is a connection between the Lax spectrum and stability spectrum.  Generally speaking, this connection could relate the eigenfunction of the linear stability problem with quadratic combinations of the eigenfunctions of the Lax problem~\cite{656,by11,bd5,bd6,bd7,bd8,t1,jm}. Therefore,  a full description of the Lax spectrum can help us to prove the spectral stability of elliptic solutions to the defocusing cmKdV equation.  For the orbital stability, the basis of our procedure is the Lyapunov method~\cite{tt11,by121,tt22,lf1,lf2,lf3}. We construct a Lyapunov functional using the conserved quantities of the defocusing cmKdV equation. Since the Lyapunov functional itself is not enough to establish orbital stability for the defocusing cmKdV equation,  our proof combines the construction of an appropriate Lyapunov functional with the seminal results of Grillakis, Shatah and Strauss~\cite{lf4}.

In Section 2, the elliptic solutions of the defocusing cmKdV equation are obtained. Then we linearize the defocusing cmKdV equation about the stationary elliptic solutions in Section 3. In Section 4, using the squared-eigenfunction connection, we conclude the spectral stability and linear stability of the elliptic solutions to the defocusing cmKdV equation. In Section 5, we introduce the NLS hierarchy and Hamiltonian structure of the flows in this hierarchy. In Section 6, We prove the orbital stability of elliptic solutions to the defocusing cmKdV equation. We present our conclusions in Section 7.
\\
\\
\noindent\textbf{\large 2. Periodic traveling-wave solutions}\\\hspace*{\parindent}
We begin by constructing the stationary solutions to defocusing case of~(\ref{cmkdv}) in the form
\begin{eqnarray}
u=e^{-i\omega t}\phi(y),\quad y=x+Vt. 
\end{eqnarray}
Then $\phi(y)$ satisfies 
\begin{eqnarray}\label{cmkdv1}
-6|\phi|^{2} \phi_{y}+\phi_{yyy}-i\omega \phi+V\phi_{y}=0.
\end{eqnarray}
Splitting $\phi$ into its amplitude and phase yields
\begin{equation}\label{ss}
\phi(y)=r(y) e^{i \theta y},
\end{equation}
where $r$ is a real-valued and bounded function of $y$, and $\theta$ is a real number.  Substituting~(\ref{ss}) into~(\ref{cmkdv1}) and separating real and imaginary parts, we have
\begin{equation}\label{tf}
\begin{array}{c}
(V-3 \theta^2) r'+r^{\prime \prime \prime}-6 r^2 r'=0,\\
r^{\prime \prime}-2 r^{3}+\left(\frac{-\theta^3+\theta V-\omega}{3\theta}\right) r=0.
\end{array}
\end{equation}
Integrating once the first equation in~(\ref{tf}), we obtain
\begin{equation}\label{tf1}
\begin{array}{c}
(V-3 \theta^2) r+r^{\prime \prime}-2 r^3=0,\\
\end{array}
\end{equation}
which implies $V-3 \theta^2=\frac{-\theta^3+\theta V-\omega}{3\theta}$. Here the integration constant is zero because of the form of the second equation of (\ref{tf}).
Multiplying~(\ref{tf1}) by $r^{\prime}$ and integrating once, we have
\begin{equation}
{r^{\prime}}^2-r^4+(V-3\theta^2)r^2+p=0,
\end{equation}
where $p$ is a constant.

%We wish to determine the condition for constructing real bounded solution $r(y)$. Equating $r^{\prime}=s$, we obtain a two-dimensional dynamical system
%\begin{equation}\label{cd}
%r^{\prime}=s, \quad s^{'}=2 r^{3}-(V-3 \theta^2) r.
%\end{equation}
%When $V-3 \theta^2>0$, we get three fixed points: $(0,0)$ and $(\pm\sqrt{\frac{(V-3 \theta^2) }{2}},0)$. After linearizing about $(0,0)$,  the resulting linear system has eigenvalues
%\begin{equation}
%\Lambda=\pm i\sqrt{V-3 \theta^2}.
%\end{equation}
%Therefore, the fixed point $(0,0)$ is a center using the linear approximation and nearby trajectories
%are closed curves. Since system~(\ref{cd}) is conservative, the fixed point is also a center
%when nonlinear terms are considered.

%After linearizing about $(\pm\sqrt{\frac{(V-3 \theta^2) }{2}},0)$,  the resulting linear system has eigenvalues
%\begin{equation}
%\Lambda=\pm\sqrt{2(V-3 \theta^2)},
%\end{equation}
%which means that $(\pm\sqrt{\frac{(V-3 \theta^2) }{2}},0)$ are saddle points.

%It is noted that periodic orbits oscillate around zero and the periodic solutions are separated from unbounded solutions by two heteroclinc orbits 
%when $V-3 \theta^2>0$. For $V-3 \theta^2<0$, we have only one fixed point $(0,0)$ which is a saddle point.  Therefore, we expect the bounded periodic solutions when $V-3 \theta^2>0$.

The Jacobi elliptic sine function with argument $y$ and
modulus $k \in[0,1)$~\cite{rb1} can be expressed as $\operatorname{sn}(y, k)$, which solves the first-order nonlinear equation
\begin{eqnarray}\label{sn1}
\left(\frac{d h}{d y}\right)^{2}=\left(1-h^{2}\right)\left(1-k^{2} h^{2}\right).
\end{eqnarray}
Motivated by~(\ref{sn1}), we obtain the Jacobi elliptic function solutions
\begin{equation}
r(y)=g \operatorname{sn}(m y, k),
\end{equation}
where 
\begin{equation}\label{re}
g^2=k^2 m^2, \quad 3\theta^2+(1+k^2)m^2-V=0,\quad p+g^2m^2=0,\quad V-3 \theta^2=\frac{-\theta^3+\theta V-\omega}{3\theta}.
\end{equation}
From~(\ref{re}), we note that $g$, $V$, $p$ and $\omega$ are expressed in terms of real-valued parameters $m$, $k$ and $\theta$. Therefore $g$, $V$, $p$ and $\omega$ are all real-valued parameters satisfying the constraints~(\ref{re}).

Here, $r(y)$ is a periodic function with period
$T(k)=\frac{4 \mathrm{K}}{m}$, where

\begin{equation}
K(k)=\int_{0}^{\pi / 2} \frac{\mathrm{d} y}{\sqrt{1-k^{2} \sin ^{2}(y)}},
\end{equation}
the complete elliptic integral of the first kind, see~\cite{rb1}.

It is noted that the nontrivial-phase solutions $\phi(y)=e^{i\theta y}r(y)$  are
quasi-periodic. However $r(y)$ is a periodic function, which will be used later.
\\
\\
\noindent\textbf{\large 3. The linear stability problem}\\\hspace*{\parindent}
To  study the orbital stability of the elliptic solutions obtained above, we consider the spectral and linear stability first. 
With the transformation 
\begin{eqnarray}
u=e^{-i\omega t}\phi(y,t),\quad y=x+Vt, 
\end{eqnarray}
Equation~(\ref{cmkdv}) could be written as 
\begin{eqnarray}\label{lcmkdv}
\phi_{\mathrm{t}}-6|\phi|^{2} \phi_{y}+\phi_{yyy}-i\omega\phi+V\phi_{y}=0.
\end{eqnarray}
Considering the perturbation of a stationary solution to (\ref{lcmkdv}),
\begin{eqnarray}\label{sp111}
\phi(y, t)=e^{i\theta y}(r(y)+\epsilon w(y, t)+i\epsilon v(y, t))+\mathcal{O}\left(\epsilon^{2}\right),
\end{eqnarray}
where $\epsilon$ is a small parameter, and $w$ and $v$ are all real-valued functions.  Substituting~(\ref{sp111}) into~(\ref{lcmkdv}), equating terms
of order $\epsilon$ and  separating real and imaginary parts, we have
\begin{equation}\label{as1}
\frac{\partial}{\partial t}\left(\begin{array}{l}
w \\
v
\end{array}\right)=J\mathcal{L}\left(\begin{array}{l}
w \\
v
\end{array}\right)=J\left(\begin{array}{cc}
L_{+} & S \\
R & L_{-}
\end{array}\right)\left(\begin{array}{l}
w \\
v
\end{array}\right),
\end{equation}
where
\begin{equation}
J=\left(\begin{array}{cc}
0 & 1 \\
-1 & 0
\end{array}\right)
\end{equation}
and linear operators $L_{+}$, $L_{-}$, $S$ and $R$ are defined by
\begin{equation}
\begin{array}{l}
L_{-}=-\theta^3+\theta V-\omega-6\theta r^2+3\theta \partial_{yy}, \\
L_{+}=-\theta^3+\theta V-\omega-18\theta r^2+3\theta \partial_{yy}, \\
S=(V-3\theta^2-6r^2) \partial_{y}+\partial_{yyy},\\
R=(-V+3\theta^2+6r^2) \partial_{y}-\partial_{yyy}+12rr^{\prime}.
\end{array}
\end{equation}
It is noted that (\ref{as1}) is  autonomous in $t$, which leads to 
\begin{equation}\label{yd1}
\left(\begin{array}{l}
w(y, t) \\
v(y, t)
\end{array}\right)=\mathrm{e}^{\lambda t}\left(\begin{array}{l}
W(y, \lambda) \\
V(y, \lambda)
\end{array}\right).
\end{equation}
Thus the spectral problem could be expressed as
\begin{equation}\label{sp1}
\lambda\left(\begin{array}{l}
W \\
V
\end{array}\right)=J\mathcal{L}\left(\begin{array}{l}
W \\
V
\end{array}\right)=J\left(\begin{array}{cc}
L_{+} & S \\
R & L_{-}
\end{array}\right)\left(\begin{array}{l}
W \\
V
\end{array}\right).
\end{equation}
We aim to prove the spectral stability of elliptic solutions analytically by determining the stability spectrum and related eigenfunctions. 
Before that, some definitions need to be introduced.
\\
\textbf{Definition 1.} \emph{The stability spectrum is the set
\begin{equation}
\sigma_{\mathcal{L}}=\left\{\lambda \in \mathbb{C}: \sup _{y \in \mathbb{R}}\left(\left|W\right|,\left|V\right|\right)<\infty\right\}.\nonumber
\end{equation}
}\emph{\textbf{Definition 2.} The solution $\phi(y)=r(y)e^{i\theta y}$  is spectrally
stable, if the spectrum $\sigma_{\mathcal{L}}$ does not intersect the open right-half of the
complex $\lambda$ plane. In particular, since (\ref{cmkdv}) is Hamiltonian, the solution $\phi(y)=r(y)e^{i\theta y}$ is spectrally
stable only if $\sigma_{\mathcal{L}}$ is a subset of the imaginary axis, i.e., $\sigma_{\mathcal{L}} \subset i\mathbb{R}$.}
\\ 
\emph{\textbf{Definition 3.} A $P$-subharmonic perturbation of a solution is a perturbation of
integer multiple $P$ times the period of the solution. }
\\
\\
%Before we determine the spectrum and related eigenfunctions of (\ref{sp1}) analytically,  we compute the spectrum numerically first with the Floquet-Fourier-Hill method~\cite{ffh1}, which is a very effective method to deal with the periodic-coefficient linear problem. 
%Since $\mathcal{L}$ has periodic coefficients, with the help of the Floquet-Bloch
%decomposition, the eigenfunctions could be expressed as 
%\begin{equation}
%\left(\begin{array}{l}
%W(y) \\
%V(y)
%\end{array}\right)=\mathrm{e}^{\mathrm{i} \mu y}\left(\begin{array}{l}
%\hat{W}(y) \\
%\hat{V}(y)
%\end{array}\right), \quad \hat{W}(y+T(k))=\hat{U}(y), \quad \hat{V}(y+T(k))=\hat{V}(y),
%\end{equation}
%with $\mu \in[-\pi / 2 T(k), \pi / 2 T(k))$.
%In Figure 1,  we show the spectral stability for (\ref{ss}) ($m=1$, $g=k=0.3$ and $\theta=0.5$) using the Floquet-Fourier-Hill method with $49$ Fourier modes and $999$ different Floquet exponents. 
%\begin{center}
%{\includegraphics[scale=0.90]
%{lv.eps}}
%\\\vspace{-0.1cm}{\footnotesize {\bf Figure~1}\
%Numerically computed spectra (imaginary part of $\lambda$ vs. real part of $\lambda$) of (\ref{ss}) with $m=1$, $g=k=0.3$ and $\theta=0.5$.}
%\end{center}

\noindent\textbf{\large 4. The Lax pair and squared eigenfunction connection}\\\hspace*{\parindent}
To show the spectral stability, we aim to construct the connection between the linear stability problem and the Lax pair.  Equation~(\ref{lcmkdv}) is an integrable equation with a Lax pair, i.e., a pair of two first-order linear ODEs
\begin{equation}\label{sjp}
\Psi_{y}=Y \Psi, \quad \Psi_{t}=T \Psi,
\end{equation}
where
\begin{equation}
Y=\left(
\begin{array}{cc}
 -i \xi & \phi(y,t) \\
 \phi^{*}(y,t) & i \xi \\
\end{array}
\right), \quad T=\left(
\begin{array}{cc}
A & B \\
 C & -A\\
\end{array}
\right),
\end{equation}
with
\begin{eqnarray}
&&A=i \xi  V-4i \xi^{3}+\frac{i \omega }{2}-2i\xi|\phi|^2+\phi_{y}\phi^{*}-\phi\phi^{*}_{y},\\
&&B=-V\phi+4\xi^2\phi+2|\phi|^2\phi+2i\xi\phi_{y}-\phi_{yy},\\
&&C=-V\phi^{*}+4\xi^2\phi^{*}+2|\phi|^2\phi^{*}-2i\xi\phi^{*}_{y}-\phi^{*}_{yy}.
\end{eqnarray}
The compatibility condition $\Psi_{yt}=\Psi_{ty}$ is equivalent to (\ref{lcmkdv}).
From the first equation of  (\ref{sjp}), one conclude that the Lax spectral problem with Lax parameter $\xi$ is self adjoint.  Therefore, the Lax spectrum is a subset of the real line
\begin{eqnarray}
\sigma_L:= \{\xi\in \mathbb{C}: \sup _{y \in \mathbb{R}}{(|\Psi_1|,|\Psi_2|)}<\infty\}\subset \mathbb{R}.\nonumber
\end{eqnarray}
Restricting to $\phi(y)=e^{i\theta y}r(y)$, we get 
\begin{equation}
T=\left(
\begin{array}{cc}
\hat{A} & \hat{B} \\
 \hat{C} & -\hat{A} \\
\end{array}
\right),
\end{equation}
where
\begin{eqnarray}
&&\hat{A}=\frac{1}{2}i \left(2V\xi-8\xi^3+\omega+4(\theta-\xi)r^2\right),\\
&&\hat{B}=e^{i \theta y} \left(r \left(\theta^2-2 \theta \xi +4 \xi ^2-V\right)-2 i (\theta-\xi ) r'-r''+2 r^3\right),\\
&&\hat{C}=e^{-i \theta y} \left(r \left(\theta^2-2 \theta \xi +4 \xi ^2-V\right)+2 i (\theta-\xi ) r'-r''+2 r^3\right).
\end{eqnarray}
Since $\hat{A}$, $\hat{B}$ and $\hat{C}$ are independent of $t$, one could write $\Psi(y,t)$ as
\begin{equation}\label{osp}
\Psi(y, t)=e^{\Omega t} \varphi(y),
\end{equation}
with $\Omega$ being independent of $t$ and $y$, which will be shown immediately.  We substitute (\ref{osp}) into $t$-part of the Lax pair and find
\begin{equation}\label{es1}
\left(\begin{array}{cc}
\hat{A}-\Omega & \hat{B} \\
\hat{C} & -\hat{A}-\Omega
\end{array}\right) \varphi=\mathbf{0}.
\end{equation}
The existence of nontrivial solutions needs
\begin{equation}\label{oc1}
\Omega^{2}=\hat{A}^{2}+\hat{B} \hat{C}=-\frac{(\theta-\xi )^2 \left(64 \theta^4 \xi ^2+128 \theta^3 \xi ^3+16 \theta^2 \left(4 \xi ^4+\xi  \omega +p\right)+16 \theta \xi ^2\omega +\omega ^2\right)}{4 \theta^2}.
\end{equation}
Using the relations~(\ref{re}), we obtain
\begin{equation}\label{oc5}
\Omega^2=-16(\xi-\xi_{5})^2(\xi-\xi_{1})(\xi-\xi_{2})(\xi-\xi_{3})(\xi-\xi_{4}),
\end{equation}
where 
\begin{eqnarray}
&&\xi_{1} =\frac{1}{2}(-\theta-m-k m),\quad \xi_{2} =\frac{1}{2}(-\theta-m+k m), \quad \xi_{3} =\frac{1}{2}(-\theta+m-k m),\nonumber\\
&&\xi_{4} =\frac{1}{2}(-\theta+m+k m),\quad \xi_{5}=\theta.
\end{eqnarray}
Here, we have determined $\Omega$ as a function of $\xi$ for $\phi(y)=e^{i\theta y}r(y)$. Then, we expect that the  eigenvector $\varphi(y)$ could be determined. In fact, from~(\ref{es1}), we have 
\begin{equation}\label{es2}
\varphi(y)=\left(\begin{array}{cc}
-\hat{B} \\
\hat{A}-\Omega
\end{array}\right) \gamma(y),
\end{equation}
where $\gamma(y)$ is a function to be determined.  With~(\ref{es2}),  we know that $\Psi(y,t)$ satisfies the $t$-part of the Lax pair. Now we substitute~(\ref{es2}) into the $x$-part of the Lax pair and obtain
\begin{equation}\label{ge1}
\gamma(y)=\gamma_{0}exp\left(\int\frac{-\hat{B}\phi^{*}+i\xi(\hat{A}-\Omega)-\hat{A}_{y}}{\hat{A}-\Omega}dy\right).
\end{equation}
It is noted that for all $\xi$ for which $\Omega\neq0$, we have constructed two linearly independent solutions of (\ref{sjp}) (one $\xi$ corresponds two different signs for $\Omega$ ). However, for $\xi$ for which $\Omega=0$, only one solution has been obtained and the second one may be constructed with reduction of order. 

In order to determine the Lax spectrum, we wish to determine for which  $\xi$,  (\ref{es2}) is bounded for all $y$. In other words, we wish to determine the set of $\xi$ such that $\gamma(y)$ is bounded. Based on~(\ref{ge1}),  we have the following necessary
and sufficient condition for boundedness
\begin{equation}\label{snc}
\left\langle\Re\left(\frac{-\hat{B}\phi^{*}+i\xi(\hat{A}-\Omega)-\hat{A}_{y}}{\hat{A}-\Omega}\right)\right\rangle=0,
\end{equation}
where $\langle\cdot\rangle=\frac{1}{T(k)} \int_{0}^{T(k)} .dy$ and $\Re$ means the real part.  Recently, Upsal and
Deconinck~\cite{jm} demonstrated that purely real Lax spectrum implies spectral stability. Our result agrees with the conclusion in~\cite{jm}. We show this explicitly below.

Since $\xi\in \mathbb{R}$, from~(\ref{oc1}) we know that $\Omega$ is real or imaginary.

$\bullet$ Case I:  For $\Omega$ being imaginary or zero,  we have 
\begin{equation}\label{snc2}
\left\langle\Re\left(\frac{-\hat{B}\phi^{*}+i\xi(\hat{A}-\Omega)-\hat{A}_{y}}{\hat{A}-\Omega}\right)\right\rangle=\frac{1}{T(k)} \int_{0}^{T(k)}\frac{2i(\theta-\xi)r^{\prime}r}{\hat{A}-\Omega} dy,
\end{equation}
which is a total derivate. Therefore  the average over a period is zero.  All $\xi$ for which $\Omega$ is imaginary are in the Lax spectrum.

$\bullet$ Case II: For $\Omega$ being real,  we have 
\begin{equation}\label{snc3}
\left\langle\Re\left(\frac{-\hat{B}\phi^{*}+i\xi(\hat{A}-\Omega)-\hat{A}_{y}}{\hat{A}-\Omega}\right)\right\rangle=\frac{1}{T(k)} \int_{0}^{T(k)}\frac{r^2(-2\theta^2-2\theta\xi+4\xi^2)}{\Omega^2+Im(\hat{A})^2}+\frac{-2i(\theta-\xi)r'\hat{A}}{\Omega^2+Im(\hat{A})^2} dy.
\end{equation}
The second term of $(\ref{snc3})$ is a total derivate, thus resulting in zero average. 
We note that 
$\left\langle\frac{r^2(-2\theta^2-2\theta\xi+4\xi^2)}{\Omega^2+Im(\hat{A})^2}\right\rangle=0$ only when $-2\theta^2-2\theta\xi+4\xi^2=0$ ($\xi=\theta$ or $\xi=-\frac{\theta}{2}$). However,
$\xi=\theta$ implies $\Omega=0$, which has been discussed in Case I. When $\xi=-\frac{\theta}{2}$, we obtain $\Omega^2=-\frac{9}{4} \theta^2 \left(k^2-1\right)^2 m^4<0$, which implies that $\Omega$ is not a real number. Therefore $\left\langle\frac{r^2(-2\theta^2-2\theta\xi+4\xi^2)}{\Omega^2+Im(\hat{A})^2}\right\rangle\neq0$ . We conclude that all $\xi$ for which $\Omega$ is real are not part of the Lax spectrum.

Based on the above analysis, we have shown that the Lax spectrum consists of all $\xi$ that makes $\Omega^2\leqslant 0$.
In order to show the set of the Lax spectrum explicitly, we need to discuss~(\ref{oc1}) and~(\ref{oc5}). 
Without loss of generality, we suppose $m>0$ (for $m<0$, we could get the similar results). We know that $\xi_{1}<\xi_{2}<\xi_{3}<\xi_{4}$. For $\xi_{5}$, we discuss the following cases:

$\bullet$ When $\theta<\frac{-(1+k)m}{3}$,  we have $\xi_{5}<\xi_{1}<\xi_{2}<\xi_{3}<\xi_{4}$, and thus the set of Lax spectrum reads (see Figure 1)
\begin{equation}\label{exam}
\sigma_{L}=\left(-\infty,\xi_{5}\right] \cup\left[\xi_{5},\xi_{1}\right] \cup\left[\xi_{2}, \xi_{3}\right] \cup\left[\xi_{4}, \infty\right).
\end{equation}
It is noted that all $\xi\in\sigma_{L}$ imply $\Omega \in i \mathbb{R}$.  Specifically, $\Omega^{2}$  takes on all negative values for $\xi\in\left(-\infty,\xi_{5}\right]$ and  $\left[\xi_{4}, \infty\right)$, which means that $\Omega$ covers the imaginary axis twice. Besides, for $\xi\in \left[\xi_{5},\xi_{1}\right]$, $\Omega^{2}$ takes on all negative values in $\left[\Omega_{1}^{2}\left(\xi^{*}\right), 0\right]$ twice, where $\Omega_{1}^{2}\left(\xi^{*}\right)$ is the local minimal value for $\xi\in \left[\xi_{5},\xi_{1}\right]$. Thus $\Omega$ covers  $\left[-i \sqrt{\left|\Omega_{1}^{2}\left(\xi^{*}\right)\right|}, i \sqrt{\left|\Omega_{1}^{2}\left(\xi^{*}\right)\right|}\right]$ twice. For $\xi\in \left[\xi_{2},\xi_{3}\right]$, $\Omega^{2}$ takes on all negative values in $\left[\Omega_{2}^{2}\left(\zeta^{*}\right), 0\right]$ twice, where $\Omega_{2}^{2}\left(\zeta^{*}\right)$ is the local minimal value for $\xi\in \left[\xi_{3},\xi_{4}\right]$. Thus $\Omega$ covers  $\left[-i \sqrt{\left|\Omega_{2}^{2}\left(\zeta^{*}\right)\right|}, i \sqrt{\left|\Omega_{2}^{2}\left(\zeta^{*}\right)\right|}\right]$ twice.
Therefore, we have 
\begin{equation}
\Omega \in(i \mathbb{R})^{2} \cup\left[-i \sqrt{\left|\Omega_{1}^{2}\left(\xi^{*}\right)\right|}, i \sqrt{\left|\Omega_{1}^{2}\left(\xi^{*}\right)\right|}\right]^{2}\cup\left[-i \sqrt{\left|\Omega_{2}^{2}\left(\zeta^{*}\right)\right|}, i \sqrt{\left|\Omega_{2}^{2}\left(\zeta^{*}\right)\right|}\right]^{2},\end{equation}
where the exponents denote multiplicities. 
\begin{center}
{\includegraphics[scale=0.90]
{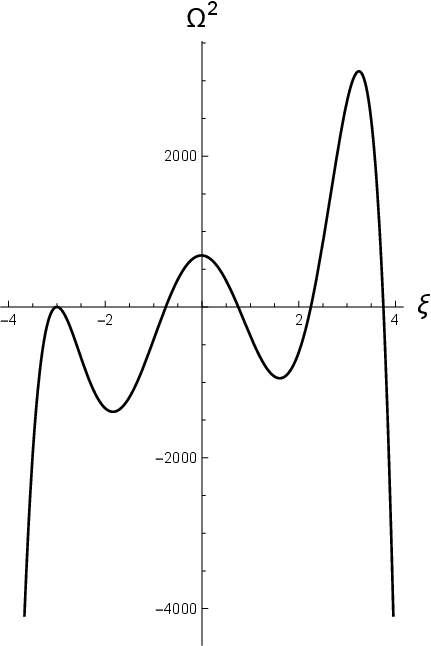}}
\\\vspace{-0.1cm}{\footnotesize {\bf Figure~1}\
Case I: $\Omega^{2}$ as a fucntion of $\xi$ with $m=3, \theta=-3$ and $k=0.5$.}
\end{center} 
$\bullet$ When $\frac{-(1+k)m}{3}<\theta<\frac{-(1-k)m}{3}$,  we have $\xi_{1}<\xi_{5}<\xi_{2}<\xi_{3}<\xi_{4}$.
Different from the first case, $\xi_{5}$ is located in $[\xi_{1}, \xi_{2}]$. 
Thus the set of Lax spectrum reads (see Figure 2)
\begin{equation}
\sigma_{L}=\left(-\infty,\xi_{1}\right] \cup\left[\xi_{2},\xi_{3}\right] \cup\left[\xi_{4}, \infty\right).
\end{equation}
Specifically, $\Omega^{2}$  takes on all negative values for $\xi\in\left(-\infty,\xi_{1}\right]$ and  $\left[\xi_{4}, \infty\right)$, which means that $\Omega$ covers the imaginary axis twice. Besides, for $\xi\in \left[\xi_{2},\xi_{3}\right]$, $\Omega^{2}$ takes on all negative values in $\left[\Omega^{2}\left(\xi^{*}\right), 0\right]$ twice, where $\Omega^{2}\left(\xi^{*}\right)$ is the local minimal value for $\xi\in \left[\xi_{2},\xi_{3}\right]$. Thus $\Omega$ covers  $\left[-i \sqrt{\left|\Omega^{2}\left(\xi^{*}\right)\right|}, i \sqrt{\left|\Omega^{2}\left(\xi^{*}\right)\right|}\right]$ twice.
Therefore, we have 
\begin{equation}
\Omega \in(i \mathbb{R})^{2} \cup\left[-i \sqrt{\left|\Omega^{2}\left(\xi^{*}\right)\right|}, i \sqrt{\left|\Omega^{2}\left(\xi^{*}\right)\right|}\right]^{2},\end{equation}
where the exponents denote multiplicities. 
\begin{center}
{\includegraphics[scale=0.90]
{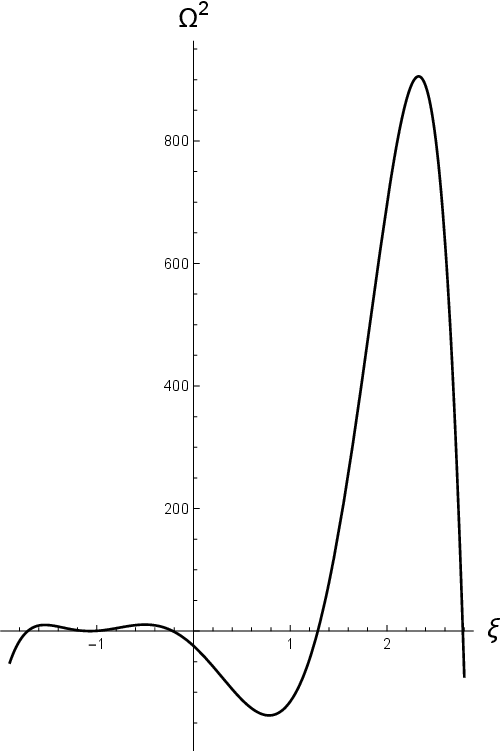}}
\\\vspace{-0.1cm}{\footnotesize {\bf Figure~2}\
Case II: $\Omega^{2}$ as a function of $\xi$ with $m=3, \theta=-1.07$ and $k=0.5$.}
\end{center}

$\bullet$ When $\frac{-(1-k)m}{3}<\theta<\frac{(1-k)m}{3}$,  we have $\xi_{1}<\xi_{2}<\xi_{5}<\xi_{3}<\xi_{4}$.
Thus the set of Lax spectrum reads (see Figure 3)
\begin{equation}
\sigma_{L}=\left(-\infty,\xi_{1}\right] \cup\left[\xi_{2},\xi_{5}\right] \cup\left[\xi_{5}, \xi_{3}\right] \cup\left[\xi_{4}, \infty\right).
\end{equation}
It is noted that all $\xi\in\sigma_{L}$ imply $\Omega \in i \mathbb{R}$.  Specifically, $\Omega^{2}$  takes on all negative values for $\xi\in\left(-\infty,\xi_{1}\right]$ and  $\left[\xi_{4}, \infty\right)$, which means that $\Omega$ covers the imaginary axis twice. Besides, for $\xi\in \left[\xi_{2},\xi_{5}\right]$, $\Omega^{2}$ takes on all negative values in $\left[\Omega_{1}^{2}\left(\xi^{*}\right), 0\right]$ twice, where $\Omega_{1}^{2}\left(\xi^{*}\right)$ is the local minimal value for $\xi\in \left[\xi_{2},\xi_{5}\right]$. Thus $\Omega$ covers  $\left[-i \sqrt{\left|\Omega_{1}^{2}\left(\xi^{*}\right)\right|}, i \sqrt{\left|\Omega_{1}^{2}\left(\xi^{*}\right)\right|}\right]$ twice. For $\xi\in \left[\xi_{5},\xi_{3}\right]$, $\Omega^{2}$ takes on all negative values in $\left[\Omega_{2}^{2}\left(\zeta^{*}\right), 0\right]$ twice, where $\Omega_{2}^{2}\left(\zeta^{*}\right)$ is the local minimal value for $\xi\in \left[\xi_{5},\xi_{3}\right]$. Thus $\Omega$ covers  $\left[-i \sqrt{\left|\Omega_{2}^{2}\left(\zeta^{*}\right)\right|}, i \sqrt{\left|\Omega_{2}^{2}\left(\zeta^{*}\right)\right|}\right]$ twice.
Therefore, we have 
\begin{equation}
\Omega \in(i \mathbb{R})^{2} \cup\left[-i \sqrt{\left|\Omega_{1}^{2}\left(\xi^{*}\right)\right|}, i \sqrt{\left|\Omega_{1}^{2}\left(\xi^{*}\right)\right|}\right]^{2}\cup\left[-i \sqrt{\left|\Omega_{2}^{2}\left(\zeta^{*}\right)\right|}, i \sqrt{\left|\Omega_{2}^{2}\left(\zeta^{*}\right)\right|}\right]^{2},\end{equation}
where the exponents denote multiplicities.
\begin{center}
{\includegraphics[scale=0.90]
{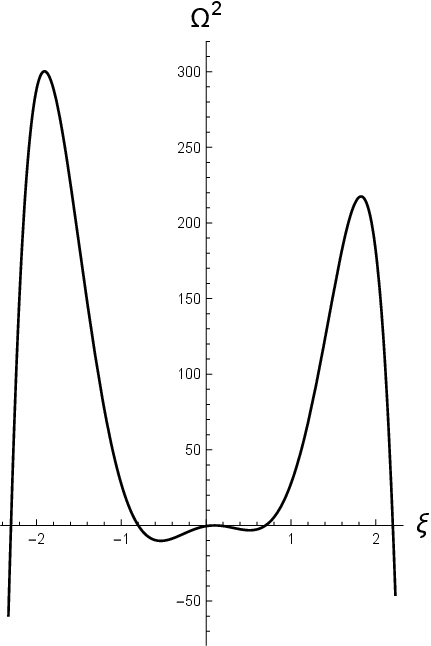}}
\\\vspace{-0.1cm}{\footnotesize {\bf Figure~3}\
Case II: $\Omega^{2}$ as a function of $\xi$ with $m=3, \theta=0.1$ and $k=0.5$.}
\end{center}
$\bullet$ When $\frac{(1-k)m}{3}<\theta<\frac{(1+k)m}{3}$,  we have $\xi_{1}<\xi_{2}<\xi_{3}<\xi_{5}<\xi_{4}$.
Thus the set of Lax spectrum reads (see Figure 4)
\begin{equation}
\sigma_{L}=\left(-\infty,\xi_{1}\right] \cup\left[\xi_{2},\xi_{3}\right] \cup\left[\xi_{4}, \infty\right).
\end{equation}
Specifically, $\Omega^{2}$  takes on all negative values for $\xi\in\left(-\infty,\xi_{1}\right]$ and  $\left[\xi_{4}, \infty\right)$, which means that $\Omega$ covers the imaginary axis twice. Besides, for $\xi\in \left[\xi_{2},\xi_{3}\right]$, $\Omega^{2}$ takes on all negative values in $\left[\Omega^{2}\left(\xi^{*}\right), 0\right]$ twice, where $\Omega^{2}\left(\xi^{*}\right)$ is the local minimal value for $\xi\in \left[\xi_{2},\xi_{3}\right]$. Thus $\Omega$ covers  $\left[-i \sqrt{\left|\Omega^{2}\left(\xi^{*}\right)\right|}, i \sqrt{\left|\Omega^{2}\left(\xi^{*}\right)\right|}\right]$ twice.
Therefore, we have 
\begin{equation}
\Omega \in(i \mathbb{R})^{2} \cup\left[-i \sqrt{\left|\Omega^{2}\left(\xi^{*}\right)\right|}, i \sqrt{\left|\Omega^{2}\left(\xi^{*}\right)\right|}\right]^{2},\end{equation}
where the exponents denote multiplicities. 
\begin{center}
{\includegraphics[scale=0.90]
{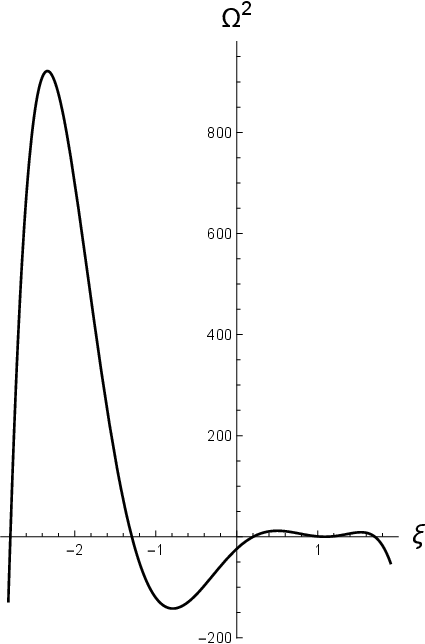}}
\\\vspace{-0.1cm}{\footnotesize {\bf Figure~4}\
Case II: $\Omega^{2}$ as a function of $\xi$ with $m=3, \theta=1.09$ and $k=0.5$.}
\end{center}

$\bullet$ When $\frac{(1+k)m}{3}<\theta$,  we have $\xi_{1}<\xi_{2}<\xi_{3}<\xi_{4}<\xi_{5}$. 

Thus the set of Lax spectrum reads (see Figure 5)
\begin{equation}
\sigma_{L}=\left(-\infty,\xi_{1}\right] \cup\left[\xi_{2},\xi_{3}\right] \cup\left[\xi_{4}, \xi_{5}\right] \cup\left[\xi_{5}, \infty\right).
\end{equation}
It is noted that all $\xi\in\sigma_{L}$ imply $\Omega \in i \mathbb{R}$.  Specifically, $\Omega^{2}$  takes on all negative values for $\xi\in\left(-\infty,\xi_{1}\right]$ and  $\left[\xi_{5}, \infty\right)$, which means that $\Omega$ covers the imaginary axis twice. Besides, for $\xi\in \left[\xi_{4},\xi_{5}\right]$, $\Omega^{2}$ takes on all negative values in $\left[\Omega_{1}^{2}\left(\xi^{*}\right), 0\right]$ twice, where $\Omega_{1}^{2}\left(\xi^{*}\right)$ is the local minimal value for $\xi\in \left[\xi_{4},\xi_{5}\right]$. Thus $\Omega$ covers  $\left[-i \sqrt{\left|\Omega_{1}^{2}\left(\xi^{*}\right)\right|}, i \sqrt{\left|\Omega_{1}^{2}\left(\xi^{*}\right)\right|}\right]$ twice. For $\xi\in \left[\xi_{2},\xi_{3}\right]$, $\Omega^{2}$ takes on all negative values in $\left[\Omega_{2}^{2}\left(\zeta^{*}\right), 0\right]$ twice, where $\Omega_{2}^{2}\left(\zeta^{*}\right)$ is the local minimal value for $\xi\in \left[\xi_{2},\xi_{3}\right]$. Thus $\Omega$ covers  $\left[-i \sqrt{\left|\Omega_{2}^{2}\left(\zeta^{*}\right)\right|}, i \sqrt{\left|\Omega_{2}^{2}\left(\zeta^{*}\right)\right|}\right]$ twice.
Therefore, we have 
\begin{equation}
\Omega \in(i \mathbb{R})^{2} \cup\left[-i \sqrt{\left|\Omega_{1}^{2}\left(\xi^{*}\right)\right|}, i \sqrt{\left|\Omega_{1}^{2}\left(\xi^{*}\right)\right|}\right]^{2}\cup\left[-i \sqrt{\left|\Omega_{2}^{2}\left(\zeta^{*}\right)\right|}, i \sqrt{\left|\Omega_{2}^{2}\left(\zeta^{*}\right)\right|}\right]^{2},\end{equation}
where the exponents denote multiplicities. 

\begin{center}
{\includegraphics[scale=0.90]
{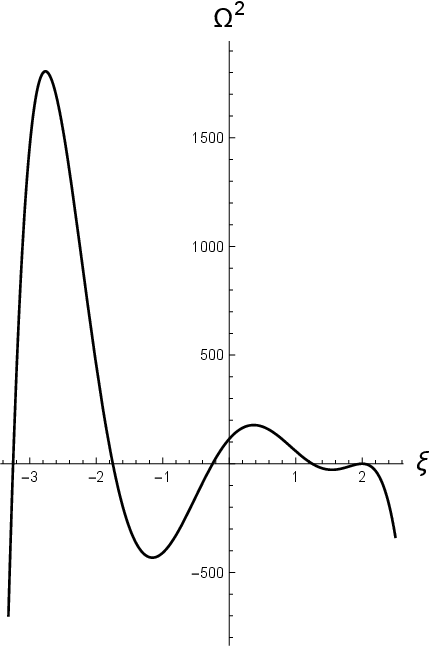}}
\\\vspace{-0.1cm}{\footnotesize {\bf Figure~5}\
Case II: $\Omega^{2}$ as a function of $\xi$ with $m=3, \theta=2$ and $k=0.5$.}
\end{center}

Many integrable systems admits the eigenfunction connections between the Lax pair and the linear stability problem~\cite{by11,bd5,bd6,bd7,bd8,t1,jm}.
To show the eigenfunction connections between the Lax pair and the linear
stability problem, the following theorem could be obtained:

\textbf{Theorem 1} \emph{The vector 
 \begin{equation}\label{yd2}
\left(w, v\right)^{T}=\left(\mathrm{e}^{-i \theta y} \Psi_{1}^{2}+\mathrm{e}^{i \theta y} \Psi_{2}^{2},-i \mathrm{e}^{-i \theta y} \Psi_{1}^{2}+i \mathrm{e}^{i \theta y} \Psi_{2}^{2}\right)^{T},
\end{equation}
satisfies the linear stability problem~(\ref{as1}). Here $\Psi=\left(\Psi_{1}, \Psi_{2}\right)^{T}$ is any solution of the Lax pair~(\ref{sjp}) with the elliptic solution $\phi=e^{i\theta y}r(y)$.
}

\textbf{Proof.}  With the linear problem~(\ref{as1}) and Lax pair~(\ref{sjp}), the proof is done by direct calculation.\ \ \ \ \ \  \ \  \  \    \    \    \  \ \ \  \              \ \   \     \     \       \     \       \         \        \ \ \ $\hfill\square$

Now we wish to build the connection between the $\sigma_{J\mathcal{L}}$ spectrum and the $\sigma_{L}$ spectrum. Substituting (\ref{yd2}) and (\ref{osp}) into (\ref{yd1}) leads to
\begin{equation}
\mathrm{e}^{\lambda t}\left(\begin{array}{l}
W \\
V
\end{array}\right)=\mathrm{e}^{2 \Omega t}\left(\begin{array}{c}
\mathrm{e}^{-i \theta y} \varphi_{1}^{2}+\mathrm{e}^{i \theta y} \varphi_{2}^{2} \\
-i \mathrm{e}^{-i \theta y} \varphi_{1}^{2}+i \mathrm{e}^{i\theta y} \varphi_{2}^{2}
\end{array}\right).
\end{equation}
Thus we obtain
\begin{equation}\label{lmd}
\lambda=2 \Omega(\xi),
\end{equation}
and
\begin{equation}\label{wt1}
\left(\begin{array}{l}
W \\
V
\end{array}\right)=\left(\begin{array}{c}
\mathrm{e}^{-i \theta y} \varphi_{1}^{2}+\mathrm{e}^{i \theta y} \varphi_{2}^{2} \\
-i \mathrm{e}^{-i \theta y} \varphi_{1}^{2}+i \mathrm{e}^{i \theta y} \varphi_{2}^{2}
\end{array}\right).
\end{equation}

 \textbf{Theorem 2} \emph{All solutions of the spectral problem (\ref{sp1}) could be derived through the squared-eigenfunction connection~(\ref{wt1}) except at $\Omega=0$}.

\textbf{Proof.}  From~(\ref{lmd}), we know that every $\lambda \in \mathbb{C}$ corresponds to one value of $\Omega$ through $\lambda=2\Omega$. The linear stability problem~(\ref{sp1}) could be viewed as a six-dimensional first-order system of ODEs.
 We define 
 \begin{equation}
 F(\xi)=\Omega^{2}-Q_{6}(\xi),
 \end{equation}
 where 
 \begin{equation}
 Q_{6}(\xi)=-\frac{(\theta-\xi )^2 \left(64 \theta^4 \xi ^2+128 \theta^3 \xi ^3+16 \theta^2 \left(4 \xi ^4+\xi  \omega +p\right)+16 \theta \xi ^2\omega +\omega ^2\right)}{4 \theta^2}. \end{equation}
 
When the discriminant of $F(\xi)$ with respect to $\xi$ does not vanish, $F(\xi)=0$ gives six values of $\xi$. Therefore, by the squared-eigenfunction connection~(\ref{wt1}), one could obtain a solution to the linear spectral problem for each of the six $\xi \in \mathbb{C}$. Now we show the six solutions generated by~(\ref{wt1}) are  linearly independent if the discriminant of $F(\xi)$ with respect to $\xi$ does not vanish.
Firstly, we rewrite $\hat{A}$ and $\hat{B}$ as
\begin{equation}\label{hahb}
\begin{aligned}
\hat{A}=\frac{1}{2} i \left(4 (\theta-\xi) r(y)^2-8 \xi ^3+2 \xi  V+\omega \right),\\
\hat{B}=-2 (\theta-\xi ) e^{i \theta y} \left((\theta+2 \xi ) r(y)+i r'(y)\right).
\end{aligned}
\end{equation}
From~(\ref{hahb}), we have 
\begin{equation}
\hat{B}_{y}=2(-i \xi \hat{B}-e^{i\theta y}r(y) \hat{A}).
\end{equation}
Besides, we rewrite~(\ref{ge1}) as 
\begin{equation}
\gamma(y)=\gamma_{0} \exp \left(-\int \frac{(\hat{A}-\Omega) \phi+\hat{B}_{y}+i \xi \hat{B}}{\hat{B}} \mathrm{~d} y\right).
\end{equation}
Then the eigenfunctions~(\ref{osp}) are written as
\begin{equation}\label{ospr}
\begin{aligned}
\Psi(y, t) &=e^{\Omega t}\left(\begin{array}{c}
-\hat{B} \\
\hat{A}-\Omega
\end{array}\right) \gamma_{0} \exp \left(-\int\left(\frac{\hat{B}_{y}}{2 \hat{B}}-\frac{e^{i\theta y}r(y)\Omega}{\hat{B}}\right) \mathrm{d} y\right) \\
&=e^{\Omega t}\left(\begin{array}{c}
-\hat{B} \\
\hat{A}-\Omega
\end{array}\right) \frac{\gamma_{0}}{\hat{B}^{1 / 2}} \exp \left(\int \frac{e^{i\theta y}r(y) \Omega}{\hat{B}} \mathrm{~d} y\right).
\end{aligned}
\end{equation}
It is noted that~(\ref{ospr}) implies that the six eigenfunctions have different singularities with different $\xi$. Therefore,  the six solutions generated by~(\ref{wt1}) are  linearly independent if the discriminant of $F(\xi)$ with respect to $\xi$ does not vanish.

When $\Omega=0$,  only one  bounded eigenfunction is obtained  through the squared-eigenfunction connection~(\ref{wt1}).

Now, we study the case that the discriminant of $F(\xi)$ with respect to $\xi$ does not vanish. The following cases should be considered:
\begin{eqnarray}
&& 1. F(\xi)=(\xi-\xi_{11})^6,\nonumber\\
&& 2. F(\xi)=(\xi-\xi_{11})^5(\xi-\xi_{12}),\nonumber\\
&& 3. F(\xi)=(\xi-\xi_{11})^4 H_{1}(\xi),\nonumber\\
&& 4. F(\xi)=(\xi-\xi_{11})^3 H_{2}(\xi),\nonumber\\
&& 5. F(\xi)=(\xi-\xi_{11})^2 H_{3}(\xi).\nonumber
\end{eqnarray}
We note that the zeros of $F(\xi)$ come from level sets of $Q_{6}(\xi)$.  As shown in Figures~1-5, $Q_{6}(\xi)$ has five different extreme points, which means that $F'(\xi)$ should admit five different zeros. This implies that case 1-4 is not possible. 
Case 5 includes the following three forms: 
\begin{eqnarray}
&& (a)  F(\xi)=(\xi-\xi_{11})^2(\xi-\xi_{12})(\xi-\xi_{13})(\xi-\xi_{14})(\xi-\xi_{15}),\nonumber\\
&& (b)  F(\xi)=(\xi-\xi_{11})^2(\xi-\xi_{12})^2(\xi-\xi_{13})^2,\nonumber\\
&& (c)  F(\xi)=(\xi-\xi_{11})^2(\xi-\xi_{12})^2(\xi-\xi_{13})(\xi-\xi_{14}).\nonumber
\end{eqnarray}
Case (b) means that three zeros of $F(\xi)$ should be equal to three extreme points of $Q_{6}(\xi)$ and $\Omega^2$ intersects with $Q_{6}$ at only three points. 
Case (c) means that two zeros of $F(\xi)$ should be equal to two extreme points of $Q_{6}(\xi)$ and  $\Omega^2$ intersects  with $Q_{6}$ at only four points.
From the graphs of $Q_{6}(\xi)$, we know that cases (b) and (c) are not possible. 
Therefore, the discriminant can vanish only in the following case:
\begin{equation} 
F(\xi)=(\xi-\xi_{11})^2(\xi-\xi_{12})(\xi-\xi_{13})(\xi-\xi_{14})(\xi-\xi_{15}).
\end{equation}
For such a case, five linearly independent solutions are obtained. The sixth solution may be derived by reduction of order, which could introduce algebraic growth and it is not an eigenfunction.   $\hfill\square$

Based on the above considerations,  we have established the  following theorem:

\textbf{Theorem 3} \emph{The periodic traveling wave solutions of
the defocusing cmKdV equation are spectrally stable. The spectrum of their associated linear stability problem is explicitly given by $\sigma(J \mathcal{L})=i \mathbb{R}$.}

As done in section 3 of~\cite{by13} and using the SCS lemma,  we conclude that the eigenfunctions are complete in  $L_{p e r}^{2}([-N \frac{T}{2}, N \frac{T}{2}])$, for any integer $N$. Therefore the linearly stable with respect to the subharmonic perturbations is proved.\\
\\
\noindent\textbf{\large 5. Hamiltonian structure and integrability}\\\hspace*{\parindent}
In next section, we wish to show the orbital stability of the elliptic solutions to the defocusing cmKdV equation by constructing a Lyapunov functional. To construct such Lyapunov functional, we need the higher-order conserved quantities of the defocusing cmKdV equation. Therefore in this section, we recall the integrability of the defocusing cmKdV equation. More importantly, we 
need rewrite the defocusing cmKdV and its Hamiltonian as a new form, which ensure that we could prove the orbital stability in next section.
Firstly, we recall the Hamiltonian structure of the complex modified KdV equation, which reads
\begin{equation}
\frac{\partial}{\partial t}\left(\begin{array}{c}
u \\
i u^{*}
\end{array}\right)=J H^{\prime}\left(u, i u^{*}\right)=J\left(\begin{array}{c}
\delta H / \delta u \\
\delta H / \delta\left(i u^{*}\right)
\end{array}\right),
\end{equation}
where 
\begin{equation}
J=\left(\begin{array}{cc}
0 & 1 \\
-1 & 0
\end{array}\right),\quad H=i \int\left(u_{x}^{*} u_{x x}+3|u|^{2} u^{*} u_{x}\right) \mathrm{d} x.
\end{equation}
The variational gradient of a function $H(u,iu)$ is defined by
\begin{equation}
H^{\prime}(u, iu^{*})=\left(\frac{\delta H}{\delta u}, \frac{\delta H}{\delta iu^{*}}\right)^{T}=\left(\sum_{j=0}^{N}(-1)^{j} \partial_{x}^{j} \frac{\partial H}{\partial u_{j x}}, \sum_{j=0}^{N}(-1)^{j} \partial_{x}^{j} \frac{\partial H}{\partial iu^{*}_{j x}}\right)^{T}.
\end{equation}
It is well known that the Hamiltonian  $H$ is one of an infinite number of conserved quantities of the NLS hierarchy.  We show some examples of the conserved quantities
\begin{equation}
\begin{aligned}
H_{0} &= \int|u|^{2} \mathrm{~d} x, \\
H_{1} &=-i \int u_{x} u^{*} \mathrm{~d} x, \\
H_{2} &=- \int\left(\left|u_{x}\right|^{2}+|u|^{4}\right) \mathrm{d} x, \\
H_{3} &=i \int\left(u_{x}^{*} u_{x x}+3|u|^{2} u^{*} u_{x}\right) \mathrm{d} x.
\end{aligned}
\end{equation}
Here, all the functionals $H_{j}$ are mutually in involution under the Poisson
bracket~\cite{sinh1,xinjia2}. The Poisson
bracket is defined as~\cite{sinh1,xinjia2}
\begin{equation}
\{H_{i}, H_{j}\}=\int \left(\begin{array}{c}
\delta H_{i} / \delta u \\
\delta H_{i} / \delta\left(i u^{*}\right)
\end{array}\right) \left(\begin{array}{cc}
0 & 1 \\
-1 & 0
\end{array}\right)\left(\begin{array}{c}
\delta H_{j} / \delta u \\
\delta H_{j} / \delta\left(i u^{*}\right)
\end{array}\right) dx.
\end{equation}
Every $H_{j}$ defines an evolution
equation with respect to a time variable $\tau_{j}$ by
\begin{equation}\label{gh1}
\frac{\partial}{\partial \tau_{j}}\left(\begin{array}{l}{u} \\ {iu^{*}}\end{array}\right)=J H_{j}^{\prime}(u, iu^{*}).
\end{equation}
For $j=3$, we know that $H_{3}=H$ is the Hamiltonian of the defocusing cmKdV.

To prove the orbital stability, we need to rewrite (\ref{gh1}) using the following transformation (this transformation is necessary and we will show this later):
\begin{equation}\label{trs2}
u=\frac{(p+il)}{\sqrt{2}},
iu^{*}=\frac{i(p-il)}{\sqrt{2}},
\end{equation}
where $p$ and $l$ are real functions of $y$ and $t$.

Using (\ref{trs2}), (\ref{gh1})  is rewritten as 

\begin{equation}
\frac{\partial}{\partial  \tau_{j}}\left(\begin{array}{l}
p \\
l
\end{array}\right)=J H_{j}^{\prime}(p, l)=J\left(\begin{array}{l}
\delta H_{j} / \delta p \\
\delta H_{j} / \delta l
\end{array}\right).
\end{equation}
When $j=3$, the defocusing cmKdV could be expressed as
\begin{equation}\label{xfc1}
\frac{\partial}{\partial t}\left(\begin{array}{l}
p \\
l
\end{array}\right)=J H_{3}^{\prime}(p, l)=\left(\begin{array}{l}
-p_{xxx}+3p^2p_{x}+3l^2p_{x} \\
-l_{xxx}+3p^2l_{x}+3l^2l_{x}
\end{array}\right),
\end{equation}
where 
\begin{equation}\label{dgl1}
H_{3}(p,l)=i \int\left(\frac{1}{2}(p_{x}-il_{x})(p_{xx}+il_{xx})+\frac{3}{4}(p^2+l^2)(p-il)(p_{x}+il_{x})\right) \mathrm{d} x.
\end{equation}
The first seven members of the hierarchy to the defocusing cmKdV read 
\begin{eqnarray}
&&u_{\tau_{0}}=-iu,\nonumber\\
&&u_{\tau_{1}}=-u_{y},\nonumber\\
&&u_{\tau_{2}}=-iu_{yy}+2i|u|^2u,\nonumber\\
&&u_{\tau_{3}}=6|u|^2u_{y}-u_{yyy},\nonumber\\
&&u_{\tau_{4}}=i(-u_{yyyy}+8|u|^{2} u_{yy}-6 u|u|^{4}+4 u\left|u_{y}\right|^{2}+6 u_{y}^{2} u^{*}+2 u^{2} u_{yy}^{*}),\nonumber\\
&&u_{\tau_{5}}=-u_{yyyyy}+10(|u|^2u_{yyy}+(u|u_{y}|^2)_{y}+2u^{*}u_{y}u_{yy}-3|u|^4u_{y}).\nonumber\\
&&u_{\tau_{6}}=2iu\left[9 u_{y}^{*} u_{yyy}+4 u_{y} u_{yyy}^{*}-35\left(u^{*}\right)^{2} u_{y}^{2}+11\left|u_{yy}\right|^{2}+6 u^{*} u_{yyyy}\right]\nonumber\\
&&+10i\left[u_{y}\left(5 u_{y}^{*} u_{yy}+3 u^{*} u_{yyy}\right)+2 u_{y}^{2} u_{yy}^{*}+2 u^{*} u_{yy}^{2}\right]+2iu^{2}\left[u_{yyyy}^{*}-5 u^{*}\left(6\left|u_{y}\right|^{2}+5 u^{*} u_{yy}\right)\right]\nonumber\\
&&-10i u^{3}\left[\left(u_{y}^{*}\right)^{2}+2 u^{*} u_{yy}^{*}\right]+20i u|u|^{6}-iu_{yyyyyy}.\nonumber
\end{eqnarray}
It has been known that every equation in this hierarchy is integrable and has a Lax pair~\cite{sinh1}. Besides, these equations share the same $y$-part Lax pair $\Psi_{y}=T_{1}\Psi$.
We show the first six $\tau_{j}$-part Lax pairs
\begin{eqnarray}
&&\Psi_{\tau_{0}}=T_{0}\Psi, \quad T_{0}=\left(
\begin{array}{cc}
 -\frac{i}{2} & 0 \\
 0 & \frac{i}{2} \\
\end{array}
\right), \nonumber\\
&&\Psi_{\tau_{1}}=T_{1}\Psi, \quad T_{1}=-\left(
\begin{array}{cc}
 -i \xi  & u \\
 u^{*} & i \xi  \\
\end{array}
\right),\nonumber\\
&&\Psi_{\tau_{2}}=T_{2}\Psi, \quad T_{2}=-2 \left(
\begin{array}{cc}
 -\frac{1}{2} i |u|^2-i \xi^2 & \xi u+\frac{1}{2} i u_{y} \\
 \xi u^{*}-\frac{1}{2} i u^{*}_{y} & \frac{1}{2} i |u|^2+i \xi^2 \\
\end{array}
\right),\nonumber\\
&&\Psi_{\tau_{3}}=T_{3}\Psi, \quad T_{3}=\left(
\begin{array}{cc}
 -4 i \xi^3-2 i |u|^2 \xi +u^{*} u_{y}-uu^{*}_{y} & 4 u \xi^2+2 i u_{y} \xi +2 |u|^2u-u_{yy} \\
 4 u^{*} \xi^2-2 i u^{*}_{y} \xi +2 |u|^2u^{*}-u^{*}_{yy} & 4 i \xi^3+2 i |u|^2  \xi -u^{*}u_{y}+uu^{*}_{y} \\
\end{array}
\right),\nonumber\\
&&\Psi_{\tau_{4}}=T_{4}\Psi,\quad T_{4}=8 \left(
\begin{array}{cc}
 A_{1} & A_{2} \\
 A_{3} & -A_{1} \\
\end{array}
\right),\nonumber\\
&&\Psi_{\tau_{5}}=T_{5}\Psi,\quad T_{5}=\left(
\begin{array}{cc}
 B_{1} & B_{2} \\
 B_{3} & -B_{1} \\
\end{array}
\right),\nonumber\\
&&\Psi_{\tau_{6}}=T_{6}\Psi,\quad T_{6}=\left(
\begin{array}{cc}
 D_{1} & D_{2} \\
 D_{3} & -D_{1} 
\end{array}
\right),\nonumber
\end{eqnarray}
where
\begin{eqnarray}
&&A_{1}= -\frac{1}{4} \xi  \left(uu^{*}_{y}-u_{y}u^{*}\right)-\frac{1}{8} i u_{y} u^{*}_{y}+\frac{1}{8} i \left(u_{yy}u^{*}+u u^{*}_{yy}\right)-\frac{1}{2} i \xi^2 |u|^2-\frac{3}{8} i |u|^4-i \xi^4,\nonumber\\
&&A_{2}=-\xi \left(\frac{1}{4} u_{yy}-\frac{1}{2} |u|^2u\right)+\frac{3}{4} i |u|^2u_{y}+\frac{1}{2} i \xi ^2 u_{y}-\frac{1}{8} i u_{yyy}+\xi^3 u,\nonumber\\
&&A_{3}=\xi  \left(\frac{1}{2} |u|^2u^{*}-\frac{1}{4} u^{*}_{yy}\right)-\frac{3}{4} i |u|^2 u^{*}_{y}-\frac{1}{2} i \xi^2 u^{*}_{y}+\frac{1}{8} i u^{*}_{yyy}+\xi^3 u^{*},\nonumber\\
&&B_{1}=4 \xi^2 \left(uu^{*}_{y}-u_{y}u^{*}\right)+2 i \xi  \left(u_{y} u^{*}_{y}-u_{yy} u^{*}-uu^{*}_{yy}+3 |u|^4\right)-u_{yy} u^{*}_{y}+u_{y} u^{*}_{yy}-6 |u|^2 u^{*}u_{y}\nonumber\\
&&+u_{yyy} u^{*}+6  |u|^2 u u^{*}_{y}-uu^{*}_{yyy}+8 i \xi^3 |u|^2+16 i \xi^5,\nonumber\\
&&B_{2}=4u u_{y}u^{*}_{y}+4 \xi^2 \left(u_{yy}-2 |u|^2u\right)+2 i \xi  \left(u_{yyy}-6 u_{y}|u|^2 \right)+6 u^2_{y} u^{*}+8 |u|^2 u_{yy}-8 i \xi^3 u_{y}\nonumber\\
&&-u_{yyyy}+2 u^2 u^{*}_{yy}-6 |u|^4u-16 \xi^4 u,\nonumber\\
&&B_{3}=4 u^{*} u_{y}u^{*}_{y}+2 u_{yy} {u^{*}}^2-4 \xi^2 \left(2 |u|^2u^{*}-u^{*}_{yy}\right)+2 i \xi  \left(6 |u|^2 u^{*}_{y}-u^{*}_{yyy}\right)+6 u (u^{*}_{y})^2+8 |u|^2 u^{*}_{yy}\nonumber\\
&&-6 |u|^4u^{*}+8 i \xi^3 u^{*}_{y}-u^{*}_{yyyy}-16 \xi^4 u^{*}.\nonumber
\end{eqnarray}
\begin{eqnarray}
&&D_1=-8 \xi ^3 (u^* u_y-u^*_y u)+\frac{1}{8} \xi ^2 (32 i (|u_y|^2-u^*_{yy} u)-32 i u^* u_{yy}+96 i |u|^4)\nonumber\\
		&&+\frac{1}{16} i \xi (-32 i u^*(6 u^*_y u^2+u_{yyy})+32 i (-u_yu^*_{yy}+u^*_y u_{yy}+u^*_{yyy} u)\nonumber\\
		&&+192 i|u|^2u^*u_y)+\frac{1}{32} (32 i(-u^*_y u_{yyy}+|u_{yy}|^2-u^*_{yyy} u_y-5(u^*_y)^2 u^2\nonumber\\
		&&+u^*_{yyyy} u)+32 i u^*(u_{yyyy}-10 u^*_{yy} u^2)-160 i (u^*)^2(u^2_y+2 uu_{yy})\nonumber\\
		&&+320 i |u|^6)+16 i \xi ^4 |u|^2 +32 i \xi ^6, \nonumber\\
&&D_2=\frac{1}{16} i \xi  (-128 i |u_y|^2u-64 i u^*_{yy} u^2-256 i|u|^2u_{yy}-192 iu^*u^2_y+192 i |u|^4u\nonumber\\
		&&+32 i u_{yyyy})+\frac{1}{32} (320 i u^*_{yy} u_y u+320 iu^*_y u_{yy}u+320 i|u_y|^2u_y-960 i|u|^4u_y\nonumber\\
		&&+320 i |u|^2u_{yyy}+640 i u^*u_yu_{yy}-32 i u_{yyyyy})+\frac{1}{4} i \xi ^3 (64 i|u|^2u-32 i u_{yy})\nonumber\\
		&&+\frac{1}{8} \xi ^2 (32 i u_{yyy}-192 i |u|^2u_y)-16 i \xi ^4 u_y-32 \xi ^5 u,\nonumber\\
&&D_3=\frac{1}{16} i \xi (-128 i|u_y|^2u^*-256i|u|^2 u^*_{yy}-192 i (u^*_y)^2 u+32 i u^*_{yyyy}-64 i (u^*)^2 u_{yy}\nonumber\\
		&&+192 i|u|^4u^*)+\frac{1}{32}(-320 i|u_y|^2u^*_y-320 iu^* u^*_{yy} u_y-320 i u^* u^*_yu_{yy}+960 i|u|^4 u^*_y \nonumber\\
		&&-640 i u^*_y u^*_{yy} u-320 i|u|^2u^*_{yyy}+32 i u^*_{yyyyy})+\frac{1}{4} i \xi ^3(64 i|u|^2 u^*-32 i u^*_{yy})\nonumber\\
		&&+\frac{1}{8} \xi ^2(192 i|u|^2u^*_y-32 i u^*_{yyy})+16 i \xi ^4 u^*_y-32 \xi ^5 u^*.\nonumber							
\end{eqnarray}

Since the members in the defocusing cmKdV hierarchy commute~\cite{sinh1,xinjia2},  one obtains the Hamiltonian system by using the linear combination of the above Hamiltonians.  
The $j$-th equation with evolution variable $\tau_{j}$ is defined as
\begin{eqnarray}
&&\frac{\partial}{\partial \tau_{j}}\left(\begin{array}{l}{p} \\ {l}\end{array}\right)=J \hat{H}_{j}^{\prime}(p, l),\\
&& \hat{H}_{j}=H_{j}+\sum_{i=0}^{j-1} c_{j, i} H_{i}, j \geqslant 1,
\end{eqnarray}
where the coefficients $c_{j,i}$ are constants that to be determined. It is noted that $\hat{H}_{3}=H_{3}+VH_{1}-\omega H_{0}$ is the Hamiltonian of the defocusing cmKdV equation~(\ref{lcmkdv}) in the traveling frame.
The Lax pair for the $j$-th equation is constructed:
\begin{eqnarray}
\Psi_{\tau_{j}}&=&\hat{T}_{j} \Psi=\left(\begin{array}{cc}{\hat{A}_{n}} & {\hat{B}_{n}} \\ {\hat{C}_{n}} & {-\hat{A}_{n}}\end{array}\right) \psi, \\ \hat{T}_{n} &=&T_{n}+\sum_{i=0}^{n-1} c_{n, i} T_{i}, \quad \hat{T}_{0} =T_{0}.
\end{eqnarray}
It is noted that any stationary solution of the defocusing cmKdV satisfies any higher-order flows with an appropriate choice of the coefficients $c_{j,i}$~\cite{sinh1,xinjia2}.
For example, the periodic traveling wave solution $\phi(y)=e^{i\theta y}r(y)$  is the stationary solution of the third equation in this hierarchy with $c_{3,0}=-\omega$, $c_{3,1}=V$ and $c_{3,2}=0$.  It is also a stationary solution to the sixth equation in this hierarchy with
\begin{eqnarray}\label{zytj}
&&c_{6,1}=2 c_{6,2} \theta+c_{6,3} V+8 c_{6,4} \theta^3-4 c_{6,4} \theta V+2 c_{6,5} \theta-4 c_{6,5} \theta^2 V\nonumber\\
&&+16 c_{6,5} \theta^4-c_{6,5} V^2-12 p \theta-16 \theta^3 V+6 \theta V^2,\\
&&c_{6,0}=-4 c_{6,2} \theta^2+c_{6,2}  V-8 c_{6,3}  \theta^3+2c_{6,3}  \theta V+2 c_{6,4}  p+c_{6,4}  V \left(4 \theta^2-V\right)\nonumber\\
&&-4 c_{6,5}  \theta \left(-2 p-6 \theta^2 V+8 \theta^4+V^2\right)-6 p V-32 \theta^4 V+64 \theta^6+V^3.\nonumber\end{eqnarray}
The condition~(\ref{zytj}) will be used to determine the orbital stability in next section.
\\
\\
\noindent\textbf{\large 6. Orbital stability}\\\hspace*{\parindent}
In order to  show the orbital stability of elliptic solutions, we rewrite~(\ref{lcmkdv}) as 
\begin{equation}\label{ssde1}
\frac{\partial}{\partial t}\left(\begin{array}{l}
p \\
l
\end{array}\right)=\left(\begin{array}{l}
-p_{yyy}+3p^2p_{y}+3l^2p_{y}-\omega l-Vp_{y} \\
-l_{yyy}+3p^2l_{y}+3l^2l_{y}+\omega p-Vl_{y}
\end{array}\right).
\end{equation}
Meanwhile, we should rewrite the linear stability according to~(\ref{ssde1}). Substituting the solution 
\begin{eqnarray}
\left(\begin{array}{l}{p(y,t)} \\ {l(y,t)}\end{array}\right)=\left(\begin{array}{l}{\hat{p}(y)} \\ {\hat{l}(y)}\end{array}\right)+\epsilon\left(\begin{array}{l}{w_{1}(y,t)} \\ {w_{2}(y,t)}\end{array}\right)+\mathcal{O}\left(\epsilon^{2}\right)
\end{eqnarray}
into~(\ref{ssde1}) and equating terms
of order $\epsilon$, we have 
\begin{eqnarray}\label{745}
w_{t}=J\mathcal{M}w,
\end{eqnarray}
where
\begin{eqnarray}
\mathcal{M}=\left(
\begin{array}{cc}
 -\omega-6\hat{l}_{y}\hat{p} &  \partial^{3}_{y}- (3\hat{p}^2+3\hat{l}^2-V)\partial_{y}-6\hat{l}\hat{l}_{y} \\
 -\partial^{3}_{y}+ (3\hat{p}^2+3\hat{l}^2-V)\partial_{y}+6\hat{p}\hat{p}_{y}& -\omega+6\hat{p}_{y}\hat{l}
\end{array}
\right).
\end{eqnarray}
Here $\mathcal{M}=\hat{H}^{\prime\prime}_{3}(\hat{p}, \hat{l})$.
Then by separating variables $(w_{1},w_{2})^{T}=e^{\lambda t}(W_{1}, W_{2})^{T}$, we have 
\begin{eqnarray}
\lambda (W_{1}, W_{2})^{T}=J\mathcal{M}(W_{1}, W_{2})^{T}.
\end{eqnarray}
As we have done in the previous section, the solutions of~(\ref{745}) are related to the Lax spectral problem via 
\begin{eqnarray}
\lambda=2\Omega(\xi),\quad  (W_{1}, W_{2})^{T}=(\Psi_{1}^2+\Psi^2_{2}, -i\Psi_{1}^2+i\Psi^2_{2})^{T},
\end{eqnarray}
which can be verified directly.

The invariance of the defocusing cmKdV equation is represented by the Lie group $G$. For $g\in G$, the elements of $G$ act on $u(y,t)$ according to 
$T(g)u(y,t)=e^{i\gamma}u(y+y_{0},t)$.  We introduce the following definition:

\textbf{Definition 4. }\emph{The stationary solution $u(y,t)=e^{-i\omega t}(\hat{p}(y)+i\hat{l}(y))$ is orbitally stable in $\mathbb{V}_{0,N}$ if for any given $\epsilon>0$ there exists a $\delta>0$ such that if $(p(y,0),l(y,0))^{T}\in\mathbb{V}_{0,N}$ then for all $t>0$
\begin{eqnarray}
\|(p(y,0),l(y,0))^{T}-(\hat{p}(y),\hat{l}(y))^{T}\|<\delta \Rightarrow inf_{g\in G}  \|(p(y,t),l(y,t))^{T}-T(g)(\hat{p}(y),\hat{l}(y))^{T}\|<\epsilon.\nonumber
\end{eqnarray}
}For this definition, one note is listed:

$\bullet$ To prove the orbital stability, we need the higher-order equations of the hierarchy, which means that $p(y,t)$ and $l(y,t)$ and their derivatives of up to third order are square integrable. Therefore, we consider the stability in the space of subharmonic functions of period $NT$
\begin{equation}
\mathbb{V}_{0,N}= H_{p e r}^{5}([-N \frac{T}{2}, N \frac{T}{2}]) \times H_{p e r}^{5}([-N \frac{T}{2}, N \frac{T}{2}]).
\end{equation}

In order to prove the orbital stability of the solution $(\hat{p}, \hat{l})$ in $\mathbb{V}_{0,N}$, we need a Lyapunov functional~\cite{by14,by15}, i.e., a constant of the motion $\mathcal{E}(\hat{p},\hat{l})$ for which $(\hat{p},\hat{l})$ is an unconstrained minimizer:
\begin{equation}\label{tc1}
\frac{d \mathcal{E}(\hat{p},\hat{l})}{d \tau}=0, \quad \mathcal{E}^{\prime}(\hat{p},\hat{l})=0, \quad\left\langle v, {\cal M}(\hat{p},\hat{l}) v\right\rangle> 0,\quad \forall v \in \mathbb{V}_{0},\quad  v \neq 0,
\end{equation}
where $\mathcal{E}^{\prime}(\hat{p},\hat{l})$ denotes the variational gradient of $\mathcal{E}$ and $\cal M$ is the Hessian of $\mathcal{E}$.
The existence of a Lyapunov functional leads to the formal stability. It is noted that  the two-dimensional null space of $\hat{H}_{3}^{\prime \prime}$ is spanned by $(-\hat{l}, \hat{p})^{T}$ and $(\hat{p}_{y},\hat{l}_{y})^{T}$, which means the kernel of $\hat{H}_{3}^{\prime \prime}$ on $\mathbb{V}_{0,N}$ is spanned by the generators of the symmetry group $G$ acting on $(\hat{p}, \hat{l})^{T}$.
With the help of results from Grillakis, Shatah, and Strauss~\cite{lf4,by14}, one could prove the orbital stability.
Since the defocusing cmKdV equation is an integrable Hamiltonian system, all the conserved quantities of such equation satisfy the first two conditions.  We just need to find one that satisfies the third condition.

To prove orbital stability,
we check the Krein signature $K_{3}$~\cite{lf4}, associated with $\hat{H}_{3}$:
\begin{equation}
K_{3}=\left\langle W, \mathcal{M} W\right\rangle=\int_{-N \frac{T}{2}}^{N  \frac{T}{2}} W^{*} \mathcal{M} W d y.
\end{equation}
Using the squared eigenfunction connection, $K_{3}$ could be expressed as 
\begin{eqnarray}\label{ks}
K_{3}&=&\left\langle W, \mathcal{M} W\right\rangle=8\Omega^2(\xi)\int_{-N \frac{K}{m}}^{N  \frac{K}{m}} \left(2V\xi-8\xi^3+\omega+4(\theta-\xi)r^2\right) d y\nonumber\\
&=&16N\Omega^2\frac{K}{m}(2V\xi-8\xi^3+\omega)+64N\Omega^2(\theta-\xi)m(K-E),\nonumber\\
&=&16N\Omega^2\left(\frac{K}{m}(2V\xi-8\xi^3+\omega)+4(\theta-\xi)m(K-E)\right),\nonumber\\
&=&-32N\Omega^2(\xi-\theta)(4 \xi^2 K+4 \xi \theta K+\theta^2 K+m^2 \left(-k^2K+K -2E\right)),\nonumber\\
&=&-32KN\Omega^2({\xi-\theta})\left((\theta+2\xi)^2+m^2\left({k^{\prime}}^2-\frac{2E}{K}\right)\right),
\end{eqnarray}
where $E(k)$ is the complete elliptic integral of the second kind~\cite{rb1}:
\begin{eqnarray}
E(k)=\int_{0}^{\pi / 2} \sqrt{1-k^{2} \sin ^{2} y} \ \mathrm{d} y.
\end{eqnarray}
There are two possibilities that lead to $K_{3}=0$.  The first one is that we choose $\xi$ for which $\Omega=0$. The second one is that we choose $\xi$ for which $P(\xi)=(2\theta+\xi)^2+m^2\left({k^{\prime}}^2-\frac{2E}{K}\right)=0$. Since ${k^{\prime}}^2-\frac{2E}{K}<0$, we obtain 
\begin{eqnarray}
\xi_{\pm c}=\frac{\pm\sqrt{m^2(\frac{2E}{K}-{k^{\prime}}^2)}-\theta}{2}.
\end{eqnarray}

\textbf{Lemma 1. } Sign changes of $K_{3}$ occur for $\xi=\xi_{\pm c}$, which is not in $\sigma_{L}$.

\textbf{Proof.} We need to show the two inequalities $\frac{-\theta-m-m k}{2}<\frac{-\sqrt{m^2(\frac{2E}{K}-{k^{\prime}}^2)}-\theta}{2}<\frac{-\theta-m+km}{2}$ and
$\frac{-\theta+m-m k}{2}<\frac{\sqrt{m^2(\frac{2E}{K}-{k^{\prime}}^2)}-\theta}{2}<\frac{-\theta+m+km}{2}$ hold.
By simplifying the above two inequalities, we find that proving the above two inequalities is equivalent to proving $1-k<\frac{E}{K}<{1+k}$.
Since $E(k)<K(k)$ and $\frac{E(k)}{K(k)}>\sqrt{1-k^2}$~\cite{rb1}, we get $1-k<\sqrt{1-k^2}<\frac{E(k)}{K(k)}<1<1+k$.
Therefore, we conclude  $\xi_{1}<\xi_{-c}<\xi_{2}$ and  $\xi_{3}<\xi_{+c}<\xi_{4}$. 
Since the Lax spectrum doesn't contain the intervals $(\xi_{1}, \xi_{2})$ and $(\xi_{3},\xi_{4})$ and the facts ($\xi_{1}<\xi_{-c}<\xi_{2}$ and  $\xi_{3}<\xi_{+c}<\xi_{4}$), we conclude that $\xi_{\pm c}$ is not in the Lax spectrum $\sigma_{L}$.
$\hfill\square$

Using Lemma 1, we conclude that  in $\sigma_{L}$, only $\xi$ for which $\Omega^2=0$ could lead to $K_{3}=0$. $K_{3}$ has different fixed signs  on the different components of $\sigma_{L}$.
Since $\hat{H}_{3}$ is not a Lyapunov functional, we need to use the higher-order conserved quantities to generate a Lyapunov functional.
Linearizing the $n$-th equation about the equilibrium solution $(\hat{p}, \hat{l})$, one obtains
\begin{eqnarray}
w_{t_{n}}=J \mathcal{L}_{n} w,
\end{eqnarray}
where $\mathcal{L}_{n}$ is the Hessian of $\hat{H}_{n}$ evaluated at the stationary solution.

Using the squared-eigenfunction connection with separation of variables gives
\begin{equation}\label{ttx1}
2 \Omega_{n} W(y)=J \mathcal{L}_{n} W(y),
\end{equation}
where $\Omega_{n}$ is defined through
\begin{equation}\label{tta}
\psi\left(y, t_{n}\right)=e^{\Omega_{n} t_{n}} \varphi(y).
\end{equation}
Substituting~(\ref{tta}) into the Lax pair of the $n$-th equation yields a relationship between $\Omega_{n}$ and $\xi$
\begin{equation}
\Omega_{n}^{2}(\xi)=\hat{A}_{n}^{2}+\hat{B}_{n} \hat{C}_{n}.
\end{equation}
As a direct result of Theorem 5 in~\cite{bd6}, we have 
\begin{equation}\label{bxz}
\Omega^2_{n}(\xi)=p^2_{n}(\xi)\Omega^2_{3}(\xi),
\end{equation}
where $p_{n}(\xi)$  is a polynomial of degree $n-3$ in $\xi$. Besides, the choice of the free parameters $c_{n,j}$
gives complete control over the roots of  $p_{n}(\xi)$.
In fact, the proof of~(\ref{bxz}) is almost the same as the cases in~\cite{bd6} (Theorem 5) and~\cite{bd7} (Section 4). 
When evaluated at a stationary solution of the $n$-th defocusing cmKdV equation, all the higher-order flows become linearly dependent, which would results in~(\ref{bxz}) through a standard AKNS calculation.

To find a Lyapunov functional,
we check $K_{6}$:
\begin{equation}
K_{6}=\int_{-N  \frac{T}{2}}^{N  \frac{T}{2}} W^{*} \mathcal{L}_{6} W d y=2 \Omega_{6} \int_{-N  \frac{T}{2}}^{N  \frac{T}{2}} W^{*} J^{-1} W d y=\frac{\Omega_{6}}{\Omega_{3}}\int_{-N  \frac{T}{2}}^{N  \frac{T}{2}} W^{*} \mathcal{L}_{3} W d y.
\end{equation}
Therefore, we obtain
\begin{equation}
K_{6}(\xi)=\Omega_{6}(\xi) \frac{K_{3}(\xi)}{\Omega_{3}(\xi)},
\end{equation}
and we use that $(\hat{p},\hat{l})$ are the stationary solutions of the fifth flow.
In order to calculate $K_{6}$, we also need the Lax pair
\begin{equation}
\hat{T}_{6}=T_{6}+c_{6,5} T_{5}+c_{6,4} T_{4}+c_{6,3} T_{3}+c_{6,2} T_{2}+c_{6,1} T_{1}+c_{6,0} T_{0}.
\end{equation}
Do not forget the condition~(\ref{zytj})  we obtained before.

The sixth NLS equation can be expressed as
\begin{equation}
\frac{\partial}{\partial \tau_{6}}\left(\begin{array}{l}{\hat{p}} \\ \hat{l}\end{array}\right)=J\left(H_{6}^{\prime}+c_{6,5} H_{5}^{\prime}+ c_{6,4} H_{4}^{\prime}+c_{6,3} H_{3}^{\prime}+c_{6,2}H_{2}^{\prime}+c_{6,1} H_{1}^{\prime}+c_{6,0} H_{0}^{\prime}\right)=0.
\end{equation}
A direct calculation gives
\begin{eqnarray}\label{ktw}
&&\Omega^2_{6}=\left(-c_{6,3}-2 c_{6,4} \xi +c_{6,5} \left(k^2+1\right) m^2+4 c_{6,5} \xi ^2+3 c_{6,5} \theta^2\right.\nonumber\\
&&\ \ \ \ \ \ \ \ \ \left.-2 \left(k^2+1\right) m^2 (\theta-\xi )+8 \xi ^3+6 \xi  \theta^2+2 \theta^3\right)^2\Omega^2_{3},
\end{eqnarray}
with 
\begin{eqnarray}
c_{6,2}=c_{6,4} \left(k^2+1\right) m^2+3 c_{6,4} \theta^2+2 \left(k^2+1\right) m^2 \theta (c_{6,5}-3 \theta)-\theta^3 (2 c_{6,5}+9 \theta)-\left(\left(k^4+4 k^2+1\right) m^4\right).\nonumber
\end{eqnarray}
Expression~(\ref{ktw}) implies that  $K_{6}$ has definite sign with whole ranges of choices for the constants $c_{6,5}, $ $c_{6,4}$ and $c_{6,3}$. Now we show this. 
In fact,  we have $K_{6}(\xi)=p_{6}(\xi) K_{3}(\xi)$, where $p_{6}(\xi)=-c_{6,3}-2 c_{6,4} \xi +c_{6,5} (k^2+1) m^2+4 c_{6,5} \xi ^2+3 c_{6,5} \theta^2-2 (k^2+1) m^2 (\theta-\xi )+8 \xi ^3+6 \xi  \theta^2+2 \theta^3$  is a polynomial in $\xi$ of degree $3$. Since we have total control over the roots
of $p_{6}(\xi)$, we choose the three constants  $c_{6,5}, $ $c_{6,4}$ and $c_{6,3}$,  so that $P_{6}(\xi)$  changes sign whenever the integral term in $K_{3}(\xi)$ changes sign.
This can be done since the integral term in $K_{3}$ is a polynomial in $\xi$ of degree $3$,  which results in $K_{6}(\xi)$ of definite sign on the entire Lax spectrum.

Since the above theory guarantees that $K_{6}$  has definite sign, we show how to choose $c_{6,5}, $ $c_{6,4}$ and $c_{6,3}$ using one example. The other cases are similar.
We consider the following Lax spectrum (corresponding to Figure 1) as an example
\begin{equation}\label{tst}
\sigma_{L}=\left(-\infty,\xi_{5}\right] \cup\left[\xi_{5},\xi_{1}\right] \cup\left[\xi_{2}, \xi_{3}\right] \cup\left[\xi_{4}, \infty\right).
\end{equation}
Here $\xi_{5}<\xi_{1}<\xi_{2}<\xi_{3}<\xi_{4}$ and $m>0$. For such case, $K_{3}\geqslant 0$ when $\xi\in\left(-\infty,\xi_{5}\right]$,  $K_{3}\leqslant 0$ when $\xi\in\left[\xi_{5},\xi_{1}\right]$, $K_{3}\geqslant 0$ when $\xi\in\left[\xi_{2}, \xi_{3}\right] $
$K_{3}\leqslant  0$ when $\xi\in\left[\xi_{4}, \infty\right)$. Do not forget that $\xi_{1}<\xi_{-c}<\xi_{2}$ and  $\xi_{3}<\xi_{+c}<\xi_{4}$.
To make $K_{6}$ has definite sign, we need to choose the parameters  $c_{6,5}, $ $c_{6,4}$ and $c_{6,3}$ and control the roots of $p_{6}(\xi)$.
To make that happen, we require that one root $(\xi_{a})$ of $p_{6}(\xi)$ is $\theta$, one root $(\xi_{b})$ of $p_{6}(\xi)$ satisfies $\xi_{1}<\xi_{b}<\xi_{2}$ and one root $(\xi_{c})$ of $p_{6}(\xi)$ satisfies $\xi_{3}<\xi_{b}<\xi_{4}$.  If the three roots of $p_{6}(\xi)$ satisfy the above conditions, $K_{6}=p_{6}K_{3}$ has definite sign. For such case, $c_{6,5}, $ $c_{6,4}$ and $c_{6,3}$ can be taken as 
\begin{eqnarray}\label{cond1}
&&c_{6,3}=-2 c_{6,4} \theta+c_{6,5} \left(k^2+1\right) m^2+7 c_{6,5} \theta^2+16 \theta^3,\nonumber\\
&&-2 k m<c_{6,5} \leq 0,\nonumber\\
&&c_{6,5} k m-c_{6,5} m+c_{6,5} \theta+2 k^2 m^2-2 k m^2+2 m^2+6 \theta^2<c_{6,4}\nonumber\\
&&<c_{6,5} k m+c_{6,5} m+c_{6,5} \theta+2 k^2 m^2+2 k m^2+2 m^2+6 \theta^2.
\end{eqnarray}
The condition~(\ref{cond1}) implies that whatever the values of $0<k<1$, $m>0$ and $\theta$ are, we can always find $c_{6,5}, $ $c_{6,4}$ and $c_{6,3}$. Therefore $K_{6}(\xi)$ has definite sign on the entire Lax spectrum.

Now we  know that $\hat{H}_{6}$ is a Lyapunov functional for the dynamics (with respect to any of the time variables in the hierarchy) of the stationary solutions.
Thus, whenever elliptic solutions are spectrally
stable with respect to subharmonic perturbations, they are formally stable in $\mathbb{V}_{0,N}$.
Since the
infinitesimal generators of the symmetries correspond to the values of $\xi$ for which $\Omega(\xi)=0$, the kernel of the functional $\hat{H}_{6}^{\prime \prime}(\hat{p}, \hat{l})$ consists of the infinitesimal generators of the symmetries of the solution $(\hat{p},\hat{l})$. As we have proved before,  $\xi_{\pm c}$ is not in $\sigma_L$.  Thus $K_{6}(\xi)=0$ is obtained only when $\Omega=0$ for $\xi\in\sigma_L$.
Therefore, we have proved Theorem 4.
\\
\\
\textbf{Theorem 4} \textbf{(Orbital stability)}
The elliptic solutions of the defocusing cmKdV equation are
orbitally stable with respect to subharmonic perturbations in $\mathbb{V}_{0,N}, N\geq 1$.
\\
\\
\noindent\textbf{\large 7. Conclusion and future work}\\\hspace*{\parindent}

\textbf{Conclusion:}  We have proven the linear stability and nonlinear stability with respect to subharmonic perturbations for the elliptic solutions of the defocusing cmKdV equation. We have established the spectral stability of elliptic solutions by explicitly computing the spectrum and the corresponding eigenfunctions associated with their linear stability problem.
By constructing an appropriate Lyapunov functional and using the seminal results of Grillakis, Shatah and Strauss~\cite{lf4}, we have shown that the elliptic solutions of the defocusing cmKdV equation are  orbitally stable with respect to subharmonic perturbations.

\textbf{Future work:} 

a) The solutions considered in this paper are  genus-one solutions and nothing is known about the stability of higher-genus solutions of the defocusing cmKdV equation. The stability of higher-genus solutions could be studied using some results from this paper along with the method from the work of Deconinck and Nivala~\cite{bd7}.

b) In this paper, we have studied the stability problems of the defocusing cmKdV equation.  For the focusing cmKdV equation, the main difficulty in constructing the stability results is that the Lax pair defines a non-self-adjoint spectral problem, which means that the Lax spectrum is not confined to the real axis.  The stability problems of the focusing cmKdV equation could be studied using the techniques from the works of Upsal, Deconinck and Segal~\cite{by90,t1,ees1}.
\\
\\

\end{document}